\documentclass[12pt]{article}
\usepackage{epsfig}
\def\be{\begin{equation}}
\def\ee{\end{equation}}
\def\bea{\begin{eqnarray}}
\def\eea{\end{eqnarray}}
\usepackage{graphicx}% Include figure files

\catcode`\@=11
\def\lsim{\mathrel{\mathpalette\@versim<}}
\def\gsim{\mathrel{\mathpalette\@versim>}}
\def\@versim#1#2{\vcenter{\offinterlineskip
\ialign{$\m@th#1\hfil##\hfil$\crcr#2\crcr\sim\crcr } }}
\catcode`\@=12
\usepackage{axodraw}

\catcode`@=12
\begin{document}
\thispagestyle{empty}
\begin{flushright}
UCRHEP-T539\\
January 2014\
\end{flushright}
\vspace{0.6in}
\begin{center}
{\LARGE \bf Vanishing Higgs Quadratic Divergence\\ 
in the Scotogenic Model and Beyond\\}
\vspace{1.2in}
{\bf Ernest Ma\\}
\vspace{0.2in}
{\sl Department of Physics and Astronomy, University of California,\\
Riverside, California 92521, USA\\}
\end{center}
\vspace{1.2in}
\begin{abstract}\
It is shown that the inherent quadratic divergence of the Higgs mass 
renormalization of the standard model may be avoided in the well-studied 
scotogenic model of radiative neutrino mass as well as other analogous 
extensions.
\end{abstract}

\newpage
\baselineskip 24pt
In quantum field theory, the additive renormalization of $m^2$ for a scalar 
field of mass $m$ is a quadratic function of the cutoff scale $\Lambda$. 
The elegant removal of this quadratic divergence is a powerful theoretical 
argument for the existence of supersymmetric particles.  However, given the 
recent discovery of the 126 GeV particle~\cite{atlas12,cms12} at the Large 
Hadron Collider (LHC), presumably the long sought Higgs boson of the 
standard model, and the nonobservation of any hint of supersymmetry, it 
may be a good time to reexamine an alternative solution of the quadratic 
divergence problem.

It was suggested a long time ago~\cite{v81} that in the standard model of 
quarks and leptons, the condition
\begin{equation}
{3 \over 2} M_W^2 + {3 \over 4} M_Z^2 + {3 \over 4} m_H^2 = \sum_f N_f m_f^2,
\end{equation}
where $N_f=3$ for quarks and $N_f=1$ for leptons, would make the coefficient 
of the $\Lambda^2$ contribution to $m_H^2$ vanish.  This would predict 
$m_H = 316$ GeV, which we now know to be incorrect.

The same idea may be extended to the case of two Higgs 
doublets~\cite{nw94,m01,go10,dk14} where 
$\langle \phi^0_{1,2} \rangle = v_{1,2}$, 
with $v = \sqrt{v_1^2 + v_2^2} = 174$ GeV.  In that case, the vanishing of 
quadratic divergences would also depend on how $\Phi_{1,2}$ couple to the 
quarks and leptons.  In the scotogenic model of radiative neutrino 
mass~\cite{m06}, there are two scalar doublets $(\phi^+,\phi^0)$ and 
$(\eta^+,\eta^0)$, distinguished from each other by a discrete $Z_2$ 
symmetry, under which $\Phi$ is even and $\eta$ odd.  Thus only 
$\phi^0$ acquires a nonzero vacuum expectation value $v$.  This 
same discrete symmetry also prevents $\eta$ from coupling to the 
usual quarks and leptons, except for the Yukawa terms
\begin{equation}
{\cal L}_Y = h_{ij} (\nu_i \eta^0 - l_i \eta^+) N_j + H.c.,
\end{equation}
where $N_j$ are three neutral singlet Majorana fermions odd under $Z_2$. 
As a result, neutrinos obtain one-loop finite radiative Majorana masses 
as shown in Fig.~1.
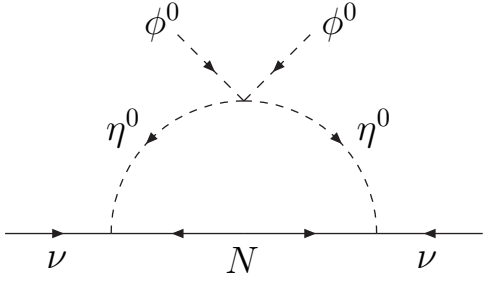
\begin{figure}[htb]
\begin{center}
\begin{picture}(360,120)(0,0)
\ArrowLine(90,10)(130,10)
\ArrowLine(180,10)(130,10)
\ArrowLine(180,10)(230,10)
\ArrowLine(270,10)(230,10)
\DashArrowLine(155,85)(180,60)3
\DashArrowLine(205,85)(180,60)3
\DashArrowArc(180,10)(50,90,180)3
\DashArrowArcn(180,10)(50,90,0)3

\Text(110,0)[]{\large $\nu$}
\Text(250,0)[]{\large $\nu$}
\Text(180,0)[]{\large $N$}
\Text(135,50)[]{\large $\eta^0$}
\Text(230,50)[]{\large $\eta^0$}
\Text(150,90)[]{\large $\phi^{0}$}
\Text(217,90)[]{\large $\phi^{0}$}

\end{picture}
\end{center}
\caption{One-loop generation of neutrino mass with $Z_2$ symmetry.}
\end{figure}
This is a well-studied model which also offers $\sqrt{2} Re(\eta^0)$ as 
a good dark-matter candidate~\cite{lnot07}.  The lightest $N$ may also 
be a dark-matter candidate~\cite{kms06} but is more suitable if the 
dark-matter discrete symmetry $Z_2$ is extended to $U(1)_D$ as proposed 
recently~\cite{mpr13}.

The scalar potential of the scotogenic $Z_2$ model is given by~\cite{m06}
\begin{eqnarray}
V &=& m_1^2 \Phi^\dagger \Phi + m_2^2 \eta^\dagger \eta + {1 \over 2} 
\lambda_1  (\Phi^\dagger \Phi)^2 + {1 \over 2} \lambda_2 (\eta^\dagger \eta)^2 
\nonumber \\ 
&+& \lambda_3  (\Phi^\dagger \Phi)(\eta^\dagger \eta) + \lambda_4 
(\Phi^\dagger \eta)(\eta^\dagger \Phi) + {1 \over 2} \lambda_5 
[(\Phi^\dagger \eta)^2 + (\eta^\dagger \Phi)^2].
\end{eqnarray}
Let $\phi^0 = v + H/\sqrt{2}$ and $\eta^0 = (\eta_R + i \eta_I)/\sqrt{2}$, then 
\begin{eqnarray}
m^2(H) &=& 2 \lambda_1 v^2, \\ 
m^2(\eta^\pm) &=& m_2^2 + \lambda_3 v^2, \\ 
m^2(\eta_R) &=& m_2^2 + (\lambda_3 + \lambda_4 + \lambda_5) v^2, \\ 
m^2(\eta_I) &=& m_2^2 + (\lambda_3 + \lambda_4 - \lambda_5) v^2. 
\end{eqnarray}
The corresponding two conditions for the vanishing of quadratic divergences 
are
\begin{eqnarray}
&& {3 \over 2} M_W^2 + {3 \over 4} M_Z^2 + {3 \over 4} m_H^2 + 
(\lambda_3 + {1 \over 2} \lambda_4) v^2 = 3 m_t^2, \\ 
&& {3 \over 2} M_W^2 + {3 \over 4} M_Z^2 + \left( {3 \over 2} \lambda_2 + 
\lambda_3 + {1 \over 2} \lambda_4 \right) v^2 = \sum_{i,j} h^2_{ij} v^2. 
\end{eqnarray}
Consequently, the following two sum rules are obtained:
\begin{eqnarray}
&& \lambda_3 + {1 \over 2} \lambda_4 = {3 \over v^2} 
\left( m_t^2 - {1 \over 2} M_W^2 - {1 \over 4} M_Z^2 - {1 \over 4} m_H^2 
\right) = 2.063, \\ 
&& h^2 - {1 \over 2} \lambda_2 = {1 \over v^2} \left( m_t^2 - {1 \over 4} 
m_H^2 \right) = 0.863,
\end{eqnarray}
where $3h^2 = \sum_{i,j} h^2_{ij}$.   Since $\lambda_2$ must be positive, 
Eq.~(11) cannot be satisfied without the Yukawa couplings of Eq.~(2).  
In other words, the existence of $N$, hence the radiative generation of 
neutrino mass, is necessary for this scenario.  In a model with simply 
a second ``inert'' scalar doublet~\cite{dm78,bhr06}, vanishing quadratic 
divergence will not be possible.  To test Eq.~(10), Eqs.~(5) to (7) may 
be used, i.e.
\begin{equation}
2 \lambda_4 v^2 = m_R^2 + m_I^2 - 2 m_+^2.
\end{equation}
As for $\lambda_3$, it may be extracted~\cite{abg12,sk13} from 
$H \to \gamma \gamma$ using also $m_+$.  However Eq.~(11) is very 
difficult to test, because $h^2$ and $\lambda_2$ are not easily 
measurable.

Analogous extensions of the scotogenic model may also accommodate vanishing 
quadratic divergences.  As an example, consider the addition of a charged 
scalar $\chi^+$ odd under $Z_2$, then the electron may acquire a radiative 
mass by assigning $e_R$ to be odd with the Yukawa couplings 
$f \bar{e}_R N_L \chi^-$ as shown in Fig.~2,
\begin{figure}[htb]
\begin{center}
\begin{picture}(360,120)(0,0)
\ArrowLine(90,10)(130,10)
\ArrowLine(130,10)(180,10)
\ArrowLine(180,10)(230,10)
\ArrowLine(230,10)(270,10)
\DashArrowLine(180,60)(180,90)3
\DashArrowArc(180,10)(50,90,180)3
\DashArrowArc(180,10)(50,0,90)3

\Text(180,10)[]{\large $\times$}
\Text(110,0)[]{\large $e_{L}$}
\Text(250,0)[]{\large $e_R$}
\Text(155,0)[]{\large $N_R$}
\Text(205,0)[]{\large $N_L$}
\Text(135,50)[]{\large $\eta^+$}
\Text(230,50)[]{\large $\chi^+$}
\Text(180,100)[]{\large $\phi^{0}$}

\end{picture}
\end{center}
\caption{One-loop generation of electron mass with soft $Z_2$ breaking.}
\end{figure}
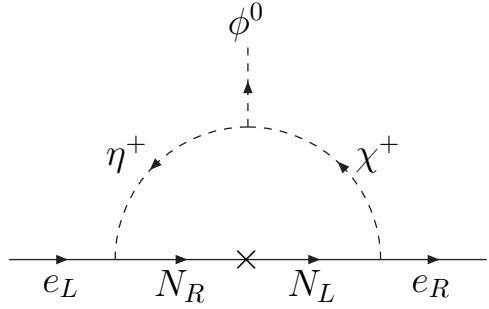
where $N_L$ is even under $Z_2$, but the soft Dirac mass term $\bar{N}_L N_R$ 
breaks $Z_2$ explicitly.  With the addition of $\chi^+$, the scalar potential 
has the extra terms
\begin{eqnarray}
V' &=& m_3^2 \chi^+ \chi^- + {1 \over 2} \lambda_6 (\chi^+ \chi^-)^2 + 
\lambda_7 (\chi^+ \chi^-)(\Phi^\dagger \Phi) + \lambda_8 (\chi^+ \chi^-) 
(\eta^\dagger \eta) \nonumber \\ 
&+& [\mu (\eta^+ \phi^0 - \eta^0 \phi^+) \chi^- + H.c.].
\end{eqnarray}
The conditions for vanishing quadratic divergence in this model are then:
\begin{eqnarray}
&& {3 \over 2} M_W^2 + {3 \over 4} M_Z^2 + {3 \over 4} m_H^2 + 
(\lambda_3 + {1 \over 2} \lambda_4 + {1 \over 2} \lambda_7) v^2 = 3 m_t^2, \\ 
&& {3 \over 2} M_W^2 + {3 \over 4} M_Z^2 + \left( {3 \over 2} \lambda_2 + 
\lambda_3 + {1 \over 2} \lambda_4 + {1 \over 2} \lambda_8 \right) v^2 
= \sum_{i,j} h^2_{ij} v^2, \\ 
&& 3 (M_Z^2 - M_W^2) + (\lambda_6 + \lambda_7 + \lambda_8) v^2 
= f^2 v^2.
\end{eqnarray}
Again, verification is possible, at least in principle.  Other more involved 
scenarios such as the scotogenic $U(1)_D$ model~\cite{mpr13} or that of a 
recent proposal~\cite{m13}, where all quark and lepton masses are radiative 
with either $Z_2$ or $U(1)_D$ dark matter, may also have similar viable solutions.

\medskip

I thank Maria Krawczyk for discussions at Scalars 2013.  This work is 
supported in part by the U.~S.~Department of Energy under Grant 
No.~DE-SC0008541.

\bibliographystyle{unsrt}

\end{document}